\documentstyle[12pt]{article}
\newcommand{\slp}{\raise.15ex\hbox{$/$}\kern-.57em\hbox{$\partial$}}
\newcommand{\sla}{\raise.15ex\hbox{$/$}\kern-.57em\hbox{$a$}}
\newcommand{\slA}{\raise.15ex\hbox{$/$}\kern-.57em\hbox{$A$}}
\newcommand{\slB}{\raise.15ex\hbox{$/$}\kern-.57em\hbox{$B$}}
\newcommand{\slb}{\raise.15ex\hbox{$/$}\kern-.57em\hbox{$b$}}
\newcommand{\slW}{\raise.15ex\hbox{$/$}\kern-.57em\hbox{$W$}}
\newcommand{\be}{\begin{equation}}
\newcommand{\ee}{\end{equation}}
\newcommand{\bear}{\begin{eqnarray}}
\newcommand{\ear}{\end{eqnarray}}

\newcommand{\D}{\cal D}

\begin{document}
\begin{flushright}
HD--THEP--95--13
\end{flushright}
\quad\\
\vspace{1.8cm}
\begin{center}
{\bf\LARGE On the Equivalence of the Maxwell-Chern-Simons Theory}\\
\medskip
{\bf\LARGE and a Self-Dual Model}\\
\vspace{1cm}
R. Banerjee\footnote[1]{On leave of absence from
S.N. Bose National Centre for Basic Sciences, DB-17, Sec 1,
Salt Lake, Calcutta-700064, India}, H. J. Rothe, and K. D. Rothe\\
\bigskip
Institut  f\"ur Theoretische Physik\\
Universit\"at Heidelberg\\
Philosophenweg 16, D-69120 Heidelberg\\
\end{center}
\vspace{2.0cm}
\begin{abstract}
We study the connection between the Green functions of the
Maxwell-Chern-Simons theory and a self-dual model by starting
from the phase-space path integral representation of the Deser-Jackiw
master Lagrangian. Their equivalence is established modulo
time-ordering ambiguities.
\end{abstract}
\newpage
In a recent interesting paper \cite{1} the bosonization of the
massive Thirring model in 2+1 dimensions was discussed by
relating it in the large mass limit to the Maxwell-Chern-Simons
(MCS) theory \cite{2}. As an intermediary step use has been made
of the equivalence \cite{3} of this theory to that of a self-dual
(SD) model discussed in \cite{4}. This analysis has been carried
out on the level of the configuration space
path-integral expressions of
the partition functions. Because of the constraint structure
associated with the various Lagrangians involved in the argument,
a complete investigation of the problem must start from a proper
phase-space path-integral formulation. This is done in the present
note. Starting from the master Lagrangian of Deser and Jackiw
\cite{3} we follow the general line of reasoning of ref. \cite{1}
and establish the equivalence, modulo time-ordering ambiguities,
of the SD model and the MCS theory on the level of Green functions.

Consider the symmetrized form of the master Lagrangian given
in \cite{3}
\be\label{1}
{\cal L}=\frac{1}{2}f^\mu f_\mu-\frac{1}{2}\epsilon^{\mu\nu\lambda}
f_\mu\partial_\nu A_\lambda-\frac{1}{2}\epsilon^{\mu\nu\lambda}A_\mu
\partial_\nu f_\lambda+\frac{m}{2}\epsilon^{\mu\nu\lambda}A_\mu
\partial_\nu A_\lambda\ee
The primary constraints \cite{5} are given by
\bear\label{2}
\Omega_0&=&\pi_0\approx0;\quad \Omega_i=\pi_i+\frac{1}{2}\epsilon
_{ij}f^j-\frac{m}{2}\epsilon_{ij}A^j\approx0\nonumber\\
\Omega_0^{(f)}&=&\pi_0^{(f)}\approx0;\quad
\Omega_i^{(f)}=\pi_i^{(f)}+\frac{1}{2}\epsilon
_{ij}A^j\approx0\ear
where $\pi_\mu(\pi_\mu^{(f)})$ are the momenta canonically conjugate
to $A^\mu(f^\mu)$. The canonical Hamiltonian is given by
\be\label{3}
H_c=\int d^2x\left[-\frac{1}{2}f^\mu f_\mu+
A_0\epsilon^{ij}\left(\partial_if_j-m\partial_iA_j\right)+\epsilon^{ij}
f_0\partial_iA_j\right]\ee
Persistency of the first-class constraints $\Omega_0$ and
$\Omega_0^{(f)}$ in time leads, respectively, to the following
secondary constraints
\bear\label{4}
\Omega_3&=&\epsilon^{ij}\partial_if_j-m\epsilon^{ij}
\partial_iA_j\approx0\nonumber\\
\Omega_3^{(f)}&=&f_0-\epsilon^{ij}\partial_iA_j\approx0\ear
Although, apart from $\pi_0$, all other constraints appear to
be second class, there actually exists a linear combination
of the constraints which is first class. This constraint
is given by
\be\label{5}
\Omega=\vec\nabla\cdot\vec\Omega+\Omega_3\approx0,\ee
which can be checked to be the generator of the gauge
transformations $A^i\to A^i+\partial^i\lambda,\
f^i\to f^i$. There are no further constraints.
Hence we have two first-class
constraints, $\Omega_0$ and $\Omega$,
and six second-class constraints
$\Omega_0^{(f)}, \Omega_i^{(f)}, \Omega_3
^{(f)}$ and $\Omega_i$. Since the equivalence to be demonstrated
refers to the observables of the SD and MCS models, we are
free to choose the Coulomb gauge for our discussion. The
phase-space partition function \cite{6} in this gauge is then
given by
\bear\label{6}
Z&=&\int Df^\mu D\pi_\mu^{(f)}D A^\mu D\pi_\mu\delta(A_0)\delta
(\vec\nabla\cdot\vec A)\delta(\Omega_0)\delta(\Omega)
\delta(\Omega_i)\nonumber\\
&&\prod^3_{\alpha=0}\delta(\Omega_\alpha^{(f)})e^{\int d^3x
\left[\pi_\mu \dot A^\mu+\pi_\mu^{(f)}\dot f^\mu-{\cal H}_c\right]}
\ear
The Faddeev-Popov determinants associated with the constraints
and the gauge-fixing are all trivial, and hence do not appear in
the functional integral. The momentum integrations in (\ref{6})
can be easily performed and one obtains
\be\label{7}
Z=\int df^\mu DA^\mu\delta(\Omega_3^{(f)})\delta(\vec\nabla\cdot
\vec A)e^{i\int d^3x{\cal L}_M}\ee
To arrive at (\ref{7}) we have expressed $\delta(\Omega_3)$ as
a Fourier integral and have redefined the $A^0$ field in order
to obtain a manifestly
Lorentz-covariant action. We next couple the
gauge-invariant fields $f^\mu$ and $F^\mu=
\epsilon^{\mu\nu\lambda}\partial_\nu
A_\lambda$ to external sources in order to establish the equivalence
of the MCS and SD models on the level of Green functions. From
(\ref{7}) we are led to consider the generating functional
\be\label{8}
Z[J,j]=\int Df^\mu DA^\mu\delta(\Omega_3^{(f)})\delta(\vec\nabla
\cdot\vec A)e^{i\int d^3x[{\cal L}+J_\mu F^\mu+
j_\mu f^\mu]}\ee
The $f_\mu$-integration is easily done to yield
\be\label{9a}
Z[J,j]=\int DA^\mu\delta(\vec\nabla\cdot\vec A)
e^{i\int d^3x[{\cal L}_{MCS}+F_\mu(J^\mu+j^\mu)+\frac{1}{2}\vec j^2]}\ee
where
\be\label{9b}
{\cal L}_{MCS}=-\frac{1}{4}F_{\mu\nu}F^{\mu\nu}+\frac{m}{2}
\epsilon^{\mu\nu\lambda}A_\mu\partial_\nu A_\lambda\ee
is the familiar MCS Lagrangian.
For vanishing sources this is the partition function of the
MCS theory in the Coulomb gauge.

Alternatively, one may perform the $A_\mu$ integration. To this
end we first integrate over the $A_0$ field, leading to
\bear\label{10a}
Z[J,j]&=&\int{\D}f^\mu{\D}A^i\delta(\vec\nabla\cdot \vec
A)\delta(\Omega_3^{(f)})\delta
(mf_0-\epsilon^{ij}\partial_if_j+\epsilon_{ij}\partial^iJ^j)
\nonumber\\
&&\times e^{i\int d^3x[{\cal L}'+f_\mu j^\mu+F_0J^0-\epsilon_{ij}
J^i\partial_0A^j]}\ear
where
\be\label{10b}
{\cal L}'=\frac{1}{2}f_\mu f^\mu-\frac{m}{2}\epsilon^{ij}A_i\partial
_0A_j-\epsilon^{ij}(f_0\partial_iA_j-f_i\partial_0
A_j)\ee
The Gaussian $A^i$ integration may be performed by expanding
the $A^i$ fields about the classical solution of the
constraint equation in the Coulomb gauge,
\be\label{11}
A^{cl}_i(\vec x,t)=\epsilon_{ij}\partial^j\int d^2\vec x'D(\vec x-
\vec{x'})f_0(\vec {x'},t)\ee
where $\vec\nabla^2(D(x-\vec{x'})=\delta(\vec x-\vec{x'})$.
One then finds that
\bear\label{12a}
Z&=&\int{\D}f_\mu\delta(mf_0-\epsilon^{ij}\partial_if_j+\epsilon
_{ij}\partial_iJ^j)\nonumber\\
&&\times\exp\left\{i\int{\cal L}_{SD}+f_\mu(J^\mu+\frac{1}{m}
\epsilon^{\mu\nu\lambda}\partial_\nu J_\lambda\right)
-\frac{1}{2m}\epsilon^{\mu\nu\lambda}J_\mu\partial_\nu
j_\lambda\ear
where
\be\label{12b}
{\cal L}_{SD}=\frac{1}{2}f_\mu f^\mu-\frac{1}{2m}\epsilon^{\mu
\lambda\nu}f_\mu\partial_\lambda f_\nu\ee
is the self-dual Lagrangian of \cite{4}.
We note that the source $J^i$ appears in
the argument of the delta-function. A more convenient
form for the computation of Green functions is obtained by
performing the integration over $f_0$. Then it can be verified
that the resulting path integral expression can also be written
in the form
\bear\label{13a}
Z[j,J]&=&\int{\cal D}f_\mu\delta(mf_0-\epsilon^{ij}
\partial_if_j)e^{i\int{\cal L}_{SD}}\nonumber\\
&\times&\exp\left\lbrace i\int\tilde f_\mu J^\mu+j^\mu f_\mu-\frac{1}{2m}
\epsilon^{\mu\nu\lambda}J_\mu\partial_\nu J_\lambda-\frac{1}
{2m^2}(\epsilon^{ij}\partial_iJ_j)^2\right.\nonumber\\
&&\left.-\frac{1}{m}j^0\epsilon^{ij}\partial_iJ_j\right\rbrace
\ear
where
\be\label{13b}
\tilde f_\mu=\frac{1}{m}\epsilon_{\mu\nu\lambda}\partial^\nu
f^\lambda\ee
is the dual of $f_\mu$.

In the absence of sources, expressions (\ref{12a}) or (\ref{13a})
reduce to the partition function associated with the SD-Lagrangian.
Recalling the alternative representation (\ref{9a}), we infer from
here the equivalence of the partition functions corresponding
to the MCS and SD models.

We next consider this equivalence on the level of Green functions.
Because of the Gaussian character of the models, it is sufficient to
consider the respective two-point functions. Functionally
differentiating  the partition functions (\ref{9a})
and (\ref{13a}) with respect to the sources $j^\mu$ and $j^\nu$,
we obtain
\be\label{14}
<F_\mu(x)F_\nu(y)>_{MCS}-i\delta_{\mu i}\delta_{\nu i}\delta(x-y)
=<f_\mu(x)f_\nu(y)>_{SD}\ee
Alternatively, by functionally differentiating (\ref{9a}) and
(\ref{13a}) with respect to the sources $J^\mu$ and $J^\nu$
one finds that
\be\label{15a}
<F_\mu(x)F_\nu(y)>_{MCS}-i\delta_{\mu i}\delta_{\nu i}\delta(x-y)=
<\tilde f_\mu(x)\tilde f_\nu(y)>_{SD}+S_{\mu\nu}(x-y)\ee
where
\be\label{15b}
S_{\mu\nu}(x-y)=-\frac{i}{m}\epsilon_{\mu\nu\lambda}
\partial^\lambda\delta(x-y)-\frac{i}{m^2}
\epsilon_{0\mu\lambda}\epsilon_{0\nu\rho}\partial^\lambda
\partial^\rho\delta(x-y)-i\delta_{\mu i}\delta_{\nu i}
\delta(x-y)\ee
Finally, by differentiating (\ref{9a}) and (\ref{13a}) with
respect to $j^\mu$ and $J^\nu$  one is led to the relation
\be\label{16a}
<F_\mu(x)F_\nu(y)>-i\delta_{\mu i}\delta_{\nu i}\delta(x-y)=
<f_\mu(x)\tilde f_\nu(y)>+S_{\mu\nu}'(x-y)\ee
where
\be\label{16b}
S_{\mu\nu}'(x-y)=-i\delta_{\mu i}\delta_{\nu i}\delta(x-y)
+\frac{i}{m}\delta_{\mu 0}\delta_{\nu j}
\epsilon_{jk}\partial_k\delta(x-y)\ee
Note that the contact term, proportional to the
$\delta$-function, contributes in the same way in all
three relations. Moreover, the remaining Schwinger-like
terms appearing in (\ref{15b}) and (\ref{16b}) can be recognized
as arising from a time-ordering ambiguity in the $\tilde f_\mu$ fields.
This can be verified by expressing these fields in terms of the
$f_\mu$'s, which are actually coupled to the sources $j^\mu$, and
making use of the commutators of the $f_\mu$ fields given in
\cite{3}. From eqs. (\ref{14}-\ref{16b})we conclude that modulo
a contact term and time-ordering ambiguities the following
identifications hold
\[ F^\mu\leftrightarrow f^\mu\leftrightarrow \tilde f^\mu\]
On the classical level this correspondence follows from the
equations of motion derived from the master Lagrangian (\ref{1})
\cite{3}. The present note therefore rounds up the analysis of refs.
\cite{3} and \cite{1}.

\medskip
\noindent{\bf Acknowledgement:} One of the authors (R.B.) would
like to thank the Alexander von Humboldt foundation for
providing the financial support which made this collaboration
possible.

\end{document}